%% file: wlcorrelator_revised.tex
\documentclass[11pt]{article}
 \pdfoutput=1
\usepackage[left=0.8in,right=0.8in,a4paper]{geometry}
\usepackage{empheq}
\usepackage{amssymb}
\usepackage[pdfstartview=FitH]{hyperref}
\usepackage{color}
\usepackage{slashed}
\usepackage{amsmath}
\usepackage{amsfonts}
\usepackage{authblk}
\usepackage{graphicx}
\usepackage{caption}
\usepackage{bbold}
\usepackage{subcaption}
\usepackage{float}
\usepackage{tikz}
\usetikzlibrary[arrows]
\usepackage{appendix}
\usepackage{listings}
\usepackage{xcolor}
\usepackage{soul}

\graphicspath{{plots/}{../plots/}}
\DeclareGraphicsExtensions{.pdf, .png, .jpg}

\definecolor{blue0}{RGB}{0,0,255}
\definecolor{blue1}{RGB}{0,51,204}
\definecolor{blue2}{RGB}{0,102,153}
\definecolor{blue3}{RGB}{0,153,102}
\definecolor{blue4}{RGB}{0,204,51}
\definecolor{blue5}{RGB}{0,255,0}

\newcommand{\be}{\begin{equation}}
\newcommand{\ee}{\end{equation}}

\def\bea{\begin{eqnarray}}
\def\eea{\end{eqnarray}}

\begin{document}

\thispagestyle{empty}

\begin{titlepage}

\begin{center}

{\Large \bf Ladder Limit for Correlators of Wilson Loops}\\

\vspace{1.5cm}

{\large \bf Diego H. Correa, Pablo Pisani, Alan Rios Fukelman}\\

\vspace{1cm}

{\it Instituto de F\'isica La Plata, CONICET, Universidad Nacional de La Plata C.C. 67, 1900}\\
{\it La Plata, Argentina}

\vspace{14pt}

\end{center}
\begin{abstract}
We study the correlator of concentric circular Wilson  loops  for arbitrary radii, spatial and internal space separations. For real values of the parameters specifying the dual string configuration, a typical Gross-Ooguri phase transition is observed. In addition, we explore some analytic continuation of a parameter $\gamma$ that characterizes the internal space separation. This enables a ladder limit in which ladder resummation and string theory computations precisely agree in the strong coupling limit. Finally, we find a critical value of $\gamma$ for which the correlator is supersymmetric and ladder diagrams can be exactly resummed for any value of the coupling constant.

\end{abstract}

\end{titlepage}

\setcounter{page}{1} \renewcommand{\thefootnote}{\arabic{footnote}}
\setcounter{footnote}{0}
\newpage

\tableofcontents

\section{Introduction}

The AdS/CFT correspondence prescribes that the computation of Wilson loops correlators can be accomplished, in the large 't Hooft coupling limit, in terms of minimal area world-sheets in AdS \cite{malda_wilson_largeN,Rey:1998ik}. In particular, a classical world-sheet stretching between two loops can be related to the connected correlation function of two Wilson loops with opposite spatial orientations \cite{Gross1998,Berenstein:1998ij}. The concomitant existence of connected and disconnected world-sheets between the two loops led to the prediction of a phase transition in the dual correlator of Wilson loops \cite{Gross1998}. This problem, for the case of concentric circular Wilson loops in ${\cal N} = 4$ super Yang-Mills has been extensively studied since the early days of the AdS/CFT correspondence proposal, either by string theory or field theory means \cite{Zarembo-correlator,Zarembo2001-breaking,Olesen:2000ji,Kim:2001td,Plefka2001,Drukker:2005cu,Burrington:2010yb,Dekel:2013kwa}. Similar problems in other alternative setups have also been considered \cite{Ahn:2006px,Armoni:2013qda,Griguolo:2012gw,Liu:2013uoa,Ziama:2015mwa,Giataganas:2015yaa,Preti:2017fhw,Aguilera-Damia:2017znn,Giombi:2018qox,Sysoeva:2018xig}.

Since generically the correlator is not supersymmetric, one cannot obtain exact results that could be taken to the strong 't Hooft coupling limit in order to compare with string theory results. The aim of this paper is to explore different regimes for the parameters characterizing the correlator of circular Wilson loops, looking for situations in which it makes sense to compare string theory with gauge theory results.

From the gauge theory perspective, the  perturbative contributions to the correlator of two Wilson loops can be classified into {\it interaction} Feynman diagrams and {\it ladder} Feynman diagrams, depending on whether they contain vertices or not. While the computation of interaction diagrams is more complicated, ladder diagrams are easier to evaluate and can even be explicitly resummed in certain cases. In particular we will consider correlators of Wilson loops with different internal space orientations parametrized by an additional parameter $\gamma$. Then, by considering an analytic continuation of this parameter we will access a regime in which ladder diagrams dominate and interaction diagrams can be dismissed. This kind of limit was originally proposed in \cite{Correa2012} for the cusp anomalous dimension and further investigated in \cite{Bykov:2012sc,Henn:2012qz,Marmiroli:2012ny,Henn:2013wfa,Bonini:2016fnc,Kim:2017sju,Cavaglia:2018lxi}.  More precisely, to access this {\it ladder limit}  we need to consider $\cos\gamma\gg1$. In this limit, to any given perturbative order $\ell$, ladder diagrams contain a factor $\cos^\ell\gamma$ while all interaction diagrams would come with lower powers of $\cos\gamma$ \cite{Correa2012}. This suggests that in this limit, the resummation of ladder diagrams will dominate the full answer and should agree, in the strong coupling limit, with the dual string theory computation.

The world-sheet dual to the correlator of concentric circular Wilson loops of radii $R_1$ and $R_2$ and separated by a distance $h$ was found in \cite{Olesen:2000ji}. In order to study the connected correlator in a ladder limit we take the Wilson loops in the correlator with different internal space orientations. As mentioned above, this introduces an additional dependence on the parameter $\gamma$ that accounts for the internal space separation. Exploring the dependence of the correlator on this additional parameter $\gamma$ will turn out to be very interesting when it comes to compare string theory with field theory computations.

Different internal space orientations for the Wilson loops in the correlator are realized, in the dual description, in terms of strings that are also extended an angle $\gamma$ in an $S^1\subset S^5$. Indeed, Drukker and Fiol found this kind of string solutions for the case of strings ending on concentric circular loops of radii $R_1$ and $R_2$ on the same plane \cite{Drukker:2005cu}. In the next section we will generalize their result for the case in which the loops are also separated by a distance $h$.

The paper is organized as follows. In section \ref{sec:string} we review the connected string configuration and generalize it for the case of arbitrary radii, spatial and internal space separations. We also study the Gross-Ooguri phase transition for the connected correlator and discuss some possible analytic extensions of the on-shell action. In section \ref{sec:fieldth} we study the contribution of ladder diagrams to the corresponding connected correlator of Wilson loops. In this way, we generalize the previous analysis of \cite{Zarembo2001-breaking} by introducing and internal space separation $\gamma$  between the two Wilson loops and analyze how the phase transition in the ladder contribution is affected by it. Finally, we study the ladder diagrams contribution for the same analytical extension of the parameters presented in the previous section and discuss its comparison with string theory results.

\section{String Solutions between concentric circles}
\label{sec:string}
Let us now consider a string whose world-sheet stretches between two concentric circles of radii $R_1$ and $R_2$ separated by a distance $h$ along a cartesian coordinate section of Euclidean $AdS_5$ and by an angle $\gamma$ in the $S^5$. Such string provides a generalization of some configurations found in previous articles \cite{Olesen:2000ji,Drukker:2005cu}. Indeed, from our set of solutions one can recover the configurations found in \cite{Olesen:2000ji} by setting $\gamma=0$. Moreover, we can also recover the configurations presented in \cite{Drukker:2005cu} by setting $h=0$, although our configuration is not strictly a generalization of this case since both settings are related by a conformal transformation.

To look for the aforementioned string configuration we consider the following Euclidean metric
\begin{equation}
ds^2 = \frac{L^2}{z^2} (dz^2 + dr^2 + r^2 d\varphi^2 + dx^2) + L^2 d\phi^2\,,
\label{eq:metrica}
\end{equation}
and make the following rotational invariant ansatz for the string embedding:
\begin{equation}
x = \sigma \,, \qquad
\varphi = \tau\,, \qquad
r = r(x)\,, \qquad
z = z(x)\,, \qquad
\phi = \phi(x) \,.
\end{equation}
The Nambu-Goto action for the string becomes\footnote{We use $\sqrt\lambda = \frac{L^2}{\alpha'}$.}
\begin{equation}
{S}_{\rm NG} = \sqrt\lambda \int dx \frac{r}{z^2}
\sqrt{1+{r'}^2 + {z'}^2 + z^2 {\phi'}^2}\,.
\label{def:lag}
\end{equation}
The fact that the Lagrangian does not depend explicitly on $\phi$ and that it is invariant under translations of $x$ gives rise to two constants of motion
\begin{equation}
\frac{r}{z^2 \sqrt{1+{r'}^2 + {z'}^2 + z^2 {\phi'}^2}} = K_x\,, \qquad
\frac{r {\phi'}}{\sqrt{1+{r'}^2 + {z'}^2 + z^2 {\phi'}^2}} = K_\phi\,,
\label{eq:constantsofmotion}
\end{equation}
while the equations of motion for the remaining variables become
\begin{equation}
\begin{split}
{r''} - \frac{r}{z^4 K_x^2} &= 0\,, \\
{z''} + \frac{2 r^2}{z^5 K_x^2} - \frac{K_\phi^2}{K_x^2 z^3} &= 0\,.
\end{split}
\label{eq:rzeoms}
\end{equation}
It is straightforward to obtain, using (\ref{eq:constantsofmotion}) and (\ref{eq:rzeoms}), the following condition
\begin{equation}
\left(r^2 + z^2 \right)'' + 2 = 0\,,
\label{eq:condicion}
\end{equation}
which, when integrated, gives
\begin{equation}
r^2 + z^2  + (x +c)^2 = a^2\,.
\label{eq:condicionintegrated}
\end{equation}
By imposing the boundary conditions $r(0)=R_1$, $r(h)=R_2$ and $z(0)=z(h)=0$, the integration constants
$a$ and $c$ are determined:
\begin{equation}
\begin{split}
c &= \frac{R_1^2-R_2^2}{2h}-\frac{h}{2}\,,
\\
a &= \sqrt{c^2+R_1^2}\,.
\end{split}
\end{equation}
As done in  \cite{Zarembo-correlator,Olesen:2000ji} Eq. (\ref{eq:condicionintegrated}) can be parametrized in terms of a trigonometric angle $\theta(x)$
\begin{equation}
\begin{split}
r &= \sqrt{a^2 - (x+c)^2} \cos\theta(x)\,,
\\
z &= \sqrt{a^2 - (x+c)^2} \sin\theta(x)\,.
\end{split}
\label{eq:cambio_var}
\end{equation}
The equations of motion further simplify to
\begin{equation}
{\theta'} = \pm \frac{a}{(a^2-(x+c)^2)}
\sqrt{\frac{\cos^2\theta - K_\phi^2 \sin^2\theta}{K_x^2 a^2 \sin^4\theta} -1}\,.
\label{eq:ecuacion_tita}
\end{equation}
The function $\theta$ grows from 0 (at $x=0$) to $\theta_0$ (at some $x=x_0$), so the $+$ sign corresponds to the interval $0\leq x\leq x_0$, while the $-$ sign corresponds to the interval $x_0\leq x \leq h$. The maximum value attained by $\theta$ satisfies
\begin{equation}
\sin^2 \theta_0 = \frac{ \sqrt{(1+K_\phi^2)^2+4a^2 K_x^2}-(1+K_\phi^2)}{2 a^2 K_x^2}\,,
\label{eq:titamax}
\end{equation}
and its position $x_0$ is simply obtained demanding $\theta' (x_0) = 0$. Eq. (\ref{eq:ecuacion_tita}) can be separated
\begin{equation}
\frac{K_x a \sin^2\theta\ d\theta}{\sqrt{\cos^2\theta - K_\phi^2 \sin^2\theta-K_x^2 a^2 \sin^4\theta}} =
\pm \frac{a dx }{(a^2-(x+c)^2)}\,,
\label{eq:ecuacion_separated}
\end{equation}
and integrated between 0 and $x_0$ and between $x_0$ and $h$ giving,
\begin{align}
\frac{1}{2} \log \left( \frac{a+x_0+c}{a-x_0-c} \right)-\frac{1}{2} \log \left( \frac{a+c}{a-c} \right) =
\int_0^{\theta_0}
d\theta \frac{K_x a \sin^2\theta}{\sqrt{\cos^2\theta - K_\phi^2 \sin^2\theta-K_x^2 a^2 \sin^4\theta}}\,,
\label{F2}
\\
\frac{1}{2} \log \left( \frac{a+h+c}{a-h-c} \right)-\frac{1}{2} \log \left( \frac{a+x_0+c}{a-x_0-c} \right) =
\int_0^{\theta_0}
d\theta \frac{K_x a \sin^2\theta}{\sqrt{\cos^2\theta - K_\phi^2 \sin^2\theta-K_x^2 a^2 \sin^4\theta}}\,.
\label{F1}
\end{align}
As we shall see, from (\ref{eq:ecuacion_tita}) and (\ref{eq:titamax}) it is possible to obtain the internal space separation $\gamma$, the on-shell action and the integrals in (\ref{F2}) and (\ref{F1}) in terms of elliptic functions. In order to simplify the expressions it is convenient to introduce a coordinate
\begin{equation}
y = \frac{\sin \theta}{\sin \theta_0}\,,
\end{equation}
and the parameters
\begin{equation}
s= \sin^2\theta_0 =\frac{ \sqrt{(1+K_\phi^2)^2+4a^2 K_x^2}-(1+K_\phi^2)}{2 a^2 K_x^2}\,,
\qquad t =  a^2 K_x^2 s^2\,.
\end{equation}
With these definitions, if we combine (\ref{F2}) and (\ref{F1}) to eliminate $x_0$, we get
\be
f(a,c,h):= \frac{1}{4} \log \left( \frac{a+c+h}{a-c-h} \right)
-\frac{1}{4} \log \left( \frac{a+c}{a-c} \right)
=
F(s,t) = \int_0^1  dy \frac{ \sqrt{s t} \ y^2}{\sqrt{1-y^2} \sqrt{1-s y^2} \sqrt{1+t y^2}}\,.
\label{fF}
\ee
We can express the internal space separation using (\ref{eq:constantsofmotion}) and (\ref{eq:cambio_var})
\begin{equation}
\gamma = \int_0^h \!dx \ \phi' =  \int_0^h dx \frac{K_\phi}{K_x}\frac{1}{[a^2 -(x+c)^2]\sin^2\theta(x)}\,,
\end{equation}
which in terms of the $y$ variable becomes
\be
\gamma= G(s,t) = \int_0^1  dy \frac{ 2\sqrt{1-s -t}}{\sqrt{1-y^2} \sqrt{1-s y^2} \sqrt{1+t y^2}}\,.
\label{Gint}
\ee

The on-shell Nambu-Goto action (\ref{eq:constantsofmotion}) gives the area of the world-sheet and is divergent as it approaches the boundary of AdS. This divergence is canceled by an appropriate boundary term \cite{Drukker:1999zq}. Equivalently,  in Poincar\'e coordinates, the regularized action can be obtained by imposing the boundary condition $z(0)=z(h)=\epsilon$, expanding for small $\epsilon$ and throwing the term order $\tfrac{1}{\epsilon}$. For the regularized on-shell action we finally get
\begin{equation}
S^{\rm reg} (s,t) = 2 \sqrt\lambda \int_0^1
\frac{dy}{y^2\sqrt{1-y^2}} \frac{1}{\sqrt{s}}
 \left(\frac{\sqrt{1-s y^2}}{\sqrt{1+t y^2}}-1\right)\,.
 \label{Sint}
\end{equation}

The integrals characterizing the classical solutions can be written in terms of elliptic functions
\begin{align}
F(s,t) =& \frac{\sqrt{t}}{\sqrt{s}}\frac{1}{\sqrt{1+t}}\left[K\left(\tfrac{s+t}{1+t}\right)-(1-s)\,\Pi\left(s\left|\tfrac{s+t}{1+t}\right.\right)\right]\,,
\label{solF}
\\
G(s,t) =& 2\frac{\sqrt{1-s-t}}{\sqrt{1+t}}K\left(\tfrac{s+t}{1+t}\right)\,,
\label{solG}
\\
S^{\rm reg} (s,t) =& -\frac{2\sqrt{\lambda}}{\sqrt{s}}\frac{1}{\sqrt{1+t}}\left[(1+t)\,E\left(\tfrac{s+t}{1+t}\right)- (1-s)\,K\left(\tfrac{s+t}{1+t}\right)\right]
\label{solS}
\end{align}

At this point, we should analyze the domain of parameters $s$ and $t$ for which one obtains real string configurations. From their definitions (\ref{eq:constantsofmotion}), the constant of motions are positive real numbers, which implies that $0\leq s\leq 1$ and $t\geq 0$. Moreover, demanding $\gamma$ to be real requires $t\leq 1-s$ as well. However, and  as motivated in the Introduction, we eventually would like to analytically continue the string configurations presented here outside the real string domain, such that $\cos\gamma$ can take any real value.

Before proceeding in this direction, let us review some aspects of the configurations in  the real string domain $0\leq s\leq 1$ and $0\leq t \leq 1-s$, thus generalizing specific cases already discussed in \cite{Zarembo-correlator,Olesen:2000ji,Drukker:2005cu,Dekel:2013kwa}.

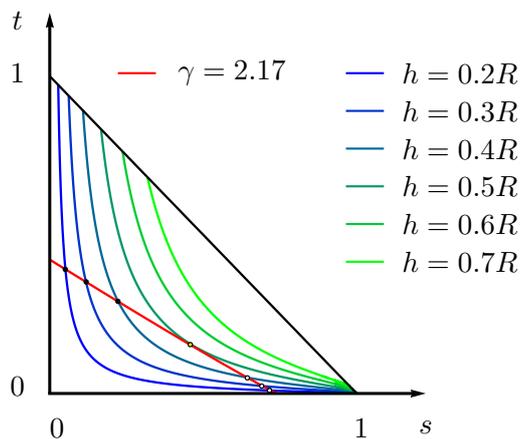
\begin{figure}[h]
\centering
\begin{tikzpicture}
\node at (-5.45,-0.85) {\includegraphics[width=5cm]{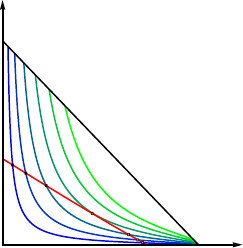}};
\node at (-2.95,-3.8) {$s$};
\node at (-3.8,-3.8) {$1$};
\node at (-7.8,-3.8) {$0$};
\node at (-8.32,1.6) {$t$};
\node at (-8.32,0.9) {$1$};
\node at (-8.32,-3.25) {$0$};
\draw[thick,red] (-7,0.9)--(-6.5,0.9);
\node at (-5.5,0.9) {$\gamma = 2.17$};
\draw[thick,blue0] (-4,0.9)--(-3.5,0.9);
\node at (-2.5,0.9) {$h = 0.2 R$};
\draw[thick,blue1] (-4,0.4)--(-3.5,0.4);
\node at (-2.5,0.4) {$h = 0.3 R$};
\draw[thick,blue2] (-4,-0.1)--(-3.5,-0.1);
\node at (-2.5,-0.1) {$h = 0.4 R$};
\draw[thick,blue3] (-4,-0.6)--(-3.5,-0.6);
\node at (-2.5,-0.6) {$h = 0.5 R$};
\draw[thick,blue4] (-4,-1.1)--(-3.5,-1.1);
\node at (-2.5,-1.1) {$h = 0.6 R$};
\draw[thick,blue5] (-4,-1.6)--(-3.5,-1.6);
\node at (-2.5,-1.6) {$h = 0.7 R$};
\end{tikzpicture}
\caption{For a given $\gamma$ and different values of $h$, black and white bullets represent stable and unstable solutions respectively. The yellow  bullet represent a critical case for which constant $\gamma$ and constant $h$ becomes tangent.}
\label{branches}
\end{figure}

To begin with, we analyze figure \ref{branches} where we plot curves for constant $\gamma$ and $h$, in the case $R_1=R_2=R$. The red curve represents some constant value for the angular separation $\gamma$ while the other curves represent different values of the spatial separation $h$, growing from blue to green tones. For the lower values of $h$ there are two intersections with the constant  $\gamma$ curve, indicating the existence of two solutions with the same values of $\gamma$ and $h$. However, as discussed in \cite{Zarembo-correlator,Olesen:2000ji,Drukker:2005cu,Dekel:2013kwa}, only one of them is stable and in our conventions it is the solution with larger $t$. As $h$ grows we reach a point where the two solutions coalesce, constituting a critical $h$ (for every value of $\gamma$) above which the connected world-sheet does no longer exist.

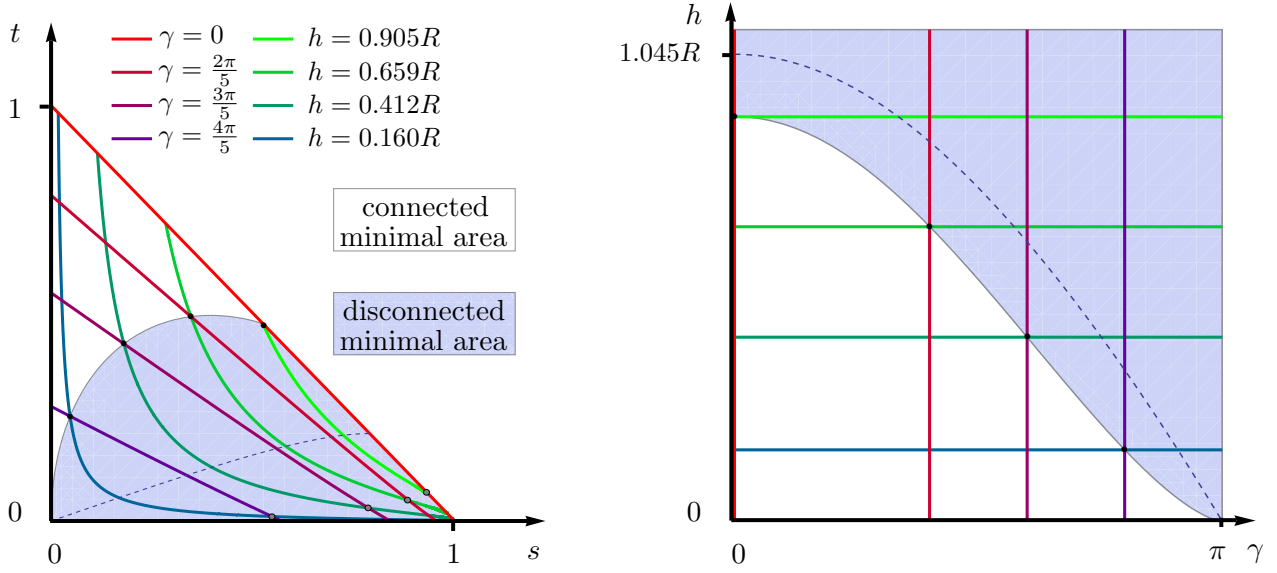
\begin{figure}[h]
\centering
\begin{tikzpicture}
\node at (0,0) {\includegraphics[width=16cm]{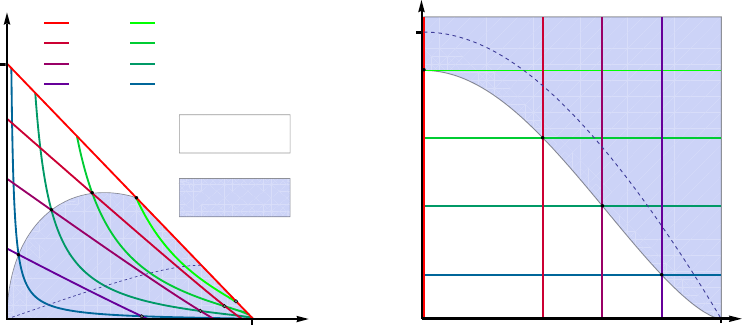}};
\node at (-1.5,-3.8) {$s$};
\node at (-2.55,-3.8) {$1$};
\node at (-7.8,-3.8) {$0$};
\node at (-8.32,3.1) {$t$};
\node at (-8.32,2.1) {$1$};
\node at (-8.32,-3.25) {$0$};
\node at (-2.94,-0.57) {disconnected};
\node at (-2.95,-0.97) {minimal area};
\node at (-2.92,0.8) {connected};
\node at (-2.95,0.4) {minimal area};
\node at (-6,3.04) {\small $\gamma=0$};
\node at (-5.9,2.60) {\small$\gamma=\tfrac{2\pi}{5}$};
\node at (-5.9,2.16) {\small $\gamma=\tfrac{3\pi}{5}$};
\node at (-5.9,1.72) {\small $\gamma=\tfrac{4\pi}{5}$};
\node at (-3.6,3.04) {\small $h=0.905R$};
\node at (-3.6,2.60) {\small$ h=0.659R$};
\node at (-3.6,2.16) {\small $h=0.412R$};
\node at (-3.6,1.72) {\small $h=0.160R$};
\node at (0.16,2.85) {\small $1.045R$};
\node at (0.61,3.35) {$h$};
\node at (0.61,-3.25) {$0$};
\node at (1.2,-3.8) {$0$};
\node at (7.5,-3.8) {$\pi$};
\node at (8.,-3.8) {$\gamma$};
\end{tikzpicture}
\caption{Connected and disconnected  world-sheet phases in the real string domain. The dashed line represents  another critical behavior. Below it no stable connected solution exist.}
\label{phases}
\end{figure}

Now, the on-shell action for every value of the angular separation $\gamma$ grows with the spatial separations $h$, so at certain critical value of $h$ the area of the connected solution becomes larger than $2\sqrt\lambda$, the value of the on-shell action of the disconnected world-sheet solution. In figure \ref{phases} we depict the connected and disconnected phases in the real string domain.  The plot in the left contains a representation of the phases in the $s$-$t$ plane. The dashed line represent the other critical behaviour: below it no stable connected solution exist.  The plot on the right represents the same but in terms of $\gamma$ and $h$.

\subsection{Analytical continuation of the string solutions}

With the possibility of implementing a ladder limit in mind, we consider certain analytical continuations of the solution (\ref{solF})-(\ref{solS}). Generically we could consider $s$ and $t$ to be complex parameters, which would correspond to complex values of the string parameters $\gamma$, $h$, $R_1$ and $R_2$. However, and for the sake of definiteness, we will restrict our analysis to values of $s$ and $t$ for which only the internal separation $\gamma$ is analytically continued, in a way such that $\cos\gamma$ can take any real value.

We shall first focus on the analytical continuation of the solution (\ref{solF})-(\ref{solS}) to the region $t>1-s$, and in particular take the limit $t\to \infty$. By expanding (\ref{solG}) we obtain
\be
\gamma =   i \log\left(\frac{16t}{1-s}\right)+{\cal O}(t^{-1}\log t)\,,
\label{expagamma}
\ee
which corresponds to
\be
\cos\gamma \approx \frac{8t}{{1-s}}\,.
\label{expacosgamma}
\ee
At this point it is already apparent that the large $t$ limit can be associated with the ladder limit motivated in the introduction. Large $t$ and $0\leq s\leq 1$ implies $\cos\gamma\gg 1$. Similarly, if we now expand (\ref{solF}) we obtain
\begin{align}
F(s,t) &= {\rm arctanh}(\sqrt s) + {\cal O}(t^{-1})\nonumber
\\
&=
\frac{1}{4} \log\left(
\frac{1+\frac{2\sqrt{s}}{1+s}}{1-\frac{2\sqrt{s}}{1+s}}\right)+{\cal O}(t^{-1})\,.
\label{expaF}
\end{align}

Since $f(a,c,h)$ can be re-written as
\be
f(a,c,h)  = \frac{1}{4}
\log\left(
\frac{1+\frac{a h}{a^2-c^2-c h}}
     {1-\frac{a h}{a^2-c^2-c h}}
     \right)\,,
\ee
one can find the relation
\be
s  = \frac{h^2+(R_1 - R_2)^2}{h^2+(R_1 + R_2)^2}\,.
\label{ssol}
\ee
Finally, we evaluate the regularized on-shell action in the limit
$t\to\infty$ and get
\be
S^{\rm reg} (s,t) = -2\sqrt\lambda \sqrt{\frac{t}{s}}+ {\cal O}(t^{-1/2})\,.
\label{expaS}
\ee
Putting together all the parameters (\ref{ssol}), (\ref{expagamma}) we obtain
\be
S^{\rm reg} (s,t)  \approx  - \sqrt\lambda e^{-\frac{i\gamma}{2}} \sqrt{\frac{R_1 R_2}{h^2+(R_1 - R_2)^2}}\,, \qquad {\rm for\ } \cos\gamma\gg 1\,.
\label{resultado0}
\ee

Thus, according to the AdS/CFT prediction this result corresponds to the strong coupling limit of the correlator of Wilson loops
\begin{equation}
\log \left(
\langle W (\mathcal{C}_1,\mathcal{C}_2)\rangle_c \right)
\simeq \sqrt{\lambda} e^{-\frac{i\gamma}{2}} \sqrt{\frac{R_1 R_2}{h^2 + (R_1-R_2)^2}}\,.
\label{resultado1}
\end{equation}
where $\mathcal{C}_1$ and $\mathcal{C}_2$ are concentric circles specified by the boundary of the world-sheet.

If we consider now the limit $t\to-\infty$, the same expansions (\ref{expagamma}), (\ref{expacosgamma}), (\ref{expaF}) and (\ref{expaS}) hold, thus corresponding to $\cos\gamma\ll -1$. However, since the action is imaginary, the disconnected world-sheet would dominate over the analytical extension in this case.

~

Let us conclude this section with a very particular case among the possible analytical continuations. By considering (\ref{Sint}) it is evident that setting $t=-s$ the regularized action becomes vanishing,
\be
S^{\rm reg} (s,-s) = 0\,.
\ee

The area of the connected world-sheet is then well above the disconnected one so the connected correlator is dominated by the latter. Nevertheless, a vanishing regularized action is as usual an interesting case to consider. Let us determine what relation implies $t=-s$ on the angular and spatial separation parameters.
For the angular separation we simply get
\be
\gamma(s,-s) = \frac{\pi}{\sqrt{1-s}}\,,
\ee
while for the function that can be related to the spatial separation and the radii we obtain
\be
F(s,-s) = \pm\frac{i}{2}\left(\pi-\gamma\right)\,.
\ee
Now using Eq. (\ref{fF}),
\be
\frac{(a+c+h)(a-c)}{(a-c-h)(a+c)} = e^{\pm 2\gamma}\,,
\ee
from which the following relation is derived
\be
\cos\gamma = -\frac{h^2 +R_1^2+R_2^2}{2 R_1 R_2}\,.
\ee
In next section it will become evident why keeping this particular relation between angular and spacial parameter is an interesting case to consider.

\section{Ladder contribution to the Wilson loop correlator}
\label{sec:fieldth}

We consider now Wilson loop operators in the fundamental representation of $U(N)$ \cite{malda_wilson_largeN,Plefka2001,Zarembo-correlator}
\begin{equation}
W(\mathcal{C}) =
\textnormal{tr}\mathcal{P} \exp
\left[\oint_{\mathcal{C}} d\tau \left(i A_\mu(x) \dot{x}^\mu + {\Phi}_i {n}^i \lvert \dot{x} \lvert  \right)\right]\,,
\label{wilsonloop}
\end{equation}
where $\mathcal{C}$ is a curve in spacetime and ${n}^i(\tau)$ an arbitrary trajectory in the internal space. We shall be interested in the connected correlator of two circular Wilson loops:
\begin{equation}
\langle W(\mathcal{C}_1,\mathcal{C}_2) \rangle_c = \langle W(\mathcal{C}_1) W(\mathcal{C}_2) \rangle - \langle W(\mathcal{C}_1) \rangle \langle W(\mathcal{C}_2) \rangle
\label{def:connected_wl}\,,
\end{equation}
where we take $\mathcal{C}_1$ and $\mathcal{C}_2$ to be concentric circles of radii $R_1$, $R_2$ respectively separated by a distance $h$, with opposite spatial orientation and different constant orientation in the internal space
\begin{equation}
\begin{split}
			\mathcal{C}_1 \,: \,x^\mu(\tau_1) &= (R_1 \cos\tau_1, R_1 \sin\tau_1,0,0 ) \, , \qquad  \ \ \, {n}^i(\tau_1) = (1,0,0,0,0,0) \, ,\\
			\mathcal{C}_2 \,: \,y^\mu(\tau_2) &= (R_2 \cos\tau_2, -R_2\sin\tau_2,h,0 ) \, , \qquad {n}^i(\tau_2) = (\cos\gamma,\sin\gamma,0,0,0,0)\,.
\end{split}
\label{def:trayect}
\end{equation}
The 1-loop contribution, {\it i.e.} the leading contribution when the 't Hooft coupling $\lambda=g^2 N\ll 1$, is simply obtained from the free propagators in the Feynman gauge
\begin{equation}
\langle \Phi_i^a(x) \Phi_j^b(y) \rangle = \frac{g^2}{4\pi^2} \frac{\delta^{ab}\delta_{ij}}{( x - y)^2} \, , \qquad \langle A_\mu^a(x) A_\nu^b(y) \rangle = \frac{g^2}{4\pi^2} \frac{\delta^{ab}\delta_{\mu \nu}}{(x-y)^2}\,.
\end{equation}

Note that the connected correlator (\ref{def:connected_wl}) only takes into account diagrams that have at least one leg in each circle, so for the trajectories (\ref{def:trayect}) we
have\footnote{We changed variable $\tau_2 \to 2\pi-\tau_2$.}
\begin{equation}
\langle W (\mathcal{C}_1,\mathcal{C}_2)\rangle_c^{\rm (1-loop)} =
\frac{\lambda}{8\pi^2}
\int_0^{2\pi} d\tau_1 \int_0^{2\pi} d\tau_2\,
\frac{1}{2}\,
\frac{\cos\gamma + \cos(\tau_1-\tau_2)}{\frac{h^2+R_1^2 + R_2^2}{2 R_1 R_2}-\cos(\tau_1-\tau_2)}\,,
\label{eq:1_loop}
\end{equation}
where the trace was taken in the fundamental representation of $U(N)$, using for the normalization of its generators ${\rm tr}(T_a T_b) = \frac{1}{2}\delta_{ab}$.

\begin{figure}[h]
\centering
\def\svgwidth{8cm}
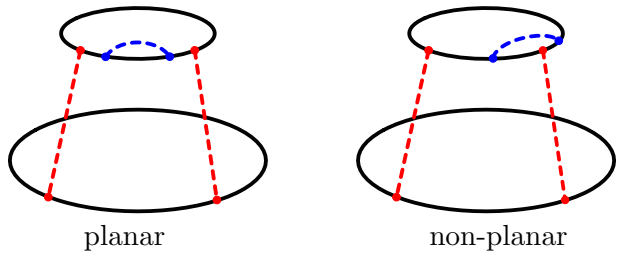
\caption{Red and blue lines represent ladder and rainbow propagators respectively. Rainbow propagators passing over the endpoints of another propagator, as in the picture on the right, render the diagram non-planar, which is suppressed in the large $N$ limit.}
\label{nonplanar}
\end{figure}

In general, to higher loop orders, contributions to the correlator can be classified into interaction and ladder diagrams, depending on whether they contain vertices or not. In this section we will be concerned with the resummation of ladder diagrams. These kind of diagrams are built exclusively with non-renormalized propagators, and we will refer to them as {\it rainbow} or {\it ladder} propagators depending on whether they extend between the same circle or they connect  $\mathcal{C}_1$ with $\mathcal{C}_2$. Each rainbow propagator behaves as in the case of the $1/2$ BPS circular Wilson loop and contributes with a constant factor to the correlator \cite{erickson_semenoff,drukker_gross_exact}. On the other hand, ladder propagators are non-trivial functions that have to be integrated. As we will see, it is convenient to organize the resummation in terms of the number of ladder propagators and to this end it is useful to define a kernel $K(\varphi)$:
\begin{equation}
K(\varphi) = \frac{1}{2}\frac{\cos\gamma + \cos(\varphi)}{\frac{h^2+R_1^2 + R_2^2}{2 R_1 R_2}-\cos(\varphi)}\,.
\label{def:kernel}
\end{equation}
We would like to retain only planar contributions. This requires that ladder propagators do not cross over each other and that rainbow propagators do not pass over the endpoints of a propagator, as for example, the diagram in the right of the figure \ref{nonplanar}.

Consider for instance a diagram with $n$ ladder propagators such that the two circles are split into $n$ arcs. Since each rainbow propagator contributes with a constant factor, the resummation of rainbows in a given arc is reduced to a matrix model problem. The resummation of rainbows in an arc of length $s$ is \cite{Zarembo2001-breaking}
\begin{equation}
W(s) = \frac{4\pi}{\sqrt\lambda s}
I_1 \left(\frac{\sqrt\lambda s}{2\pi}\right)\,,
\end{equation}
where $I_1$ is a modified Bessel function. Then, the connected correlator can be expanded as
\begin{equation}
\langle W (\mathcal{C}_1,\mathcal{C}_2)\rangle_c = \sum_{n=1}^\infty
\left(\frac{\lambda}{8\pi^2}\right)^n {\cal W}_n
\,,
\label{eq:expa ladder}
\end{equation}
where ${\cal W}_n$ is the resummation of all diagrams with $n$ ladders and is given
by\footnote{As in (\ref{eq:1_loop}), we change variables $s_a\to 2\pi-s_a$  to deal with more symmetric expressions.}
\begin{align}
{\cal W}_n = & \underset{0<t_1<t_2<\cdots<t_{n}<2\pi}{\int\! dt_1\!\int \!dt_2\!\cdots\!\int dt_n}
\underset{0<s_1<s_2<\cdots<s_{n}<2\pi}{\int \!ds_1\!\int \!ds_2\cdots\!\int ds_n}  \sum_{j=1}^{n}\prod_{a=1}^n K(t_a-s_{a+j})\label{nladder}
\\
&\times W(t_2-t_1)\cdots W(t_n-t_{n-1})W(2\pi-t_n+t_1)
W(s_2-s_1)\cdots W(s_n-s_{n-1})W(2\pi-s_n+s_1)\nonumber
 \,,
\end{align}
where we identify $s_b$ with $s_{b-n}$, whenever $b>n$.

We would like to have some exact evaluation of the resummation (\ref{eq:expa ladder}) and eventually consider it in the strong coupling limit. This problem, for the particular case $\gamma =0$, was already studied in \cite{Zarembo2001-breaking} by writing a Dyson equation which can be solved in the strong coupling limit. In principle, we could simply take this strong coupling result and replace its kernel by the $\gamma$ dependent one (\ref{def:kernel}).

However, we have found some discrepancy of the expansion (\ref{nladder}) with the expansion of the Dyson equation proposed in \cite{Zarembo2001-breaking}. We emphasize that this inconsistency does not affect the strong coupling description nor the conclusions about the phase transition made in \cite{Zarembo2001-breaking}. Nevertheless, and given that we are interested not only in the strong coupling limit, we would like to carefully account for the expansion (\ref{eq:expa ladder}).

The sum of $n$ terms in ${\cal W}_n$ is due to the existence of $n$ inequivalent planar ways of setting the
ladder propagators, but we can collect all of them into a single term by appropriate changes of coordinates.
Let us consider the following set of change of variables
\begin{equation}
t_a =
\left\{
\begin{array}{lllc}
\tilde t_n - \tilde t_{i-a}& & {\rm if}& 1\leq a\leq i-1
\\
\tilde t_n & & {\rm if}&  a= i
\\
\tilde t_n - \tilde t_{n+i-a} + 2 \pi & & {\rm if}& a > i
\end{array}
\right.
\label{changet}
\end{equation}
For $i = 1, \cdots, n$ they constitute $n$ different changes of variables. The new variables are ordered as follows
\begin{align*}
  \tilde t_1<\cdots<\tilde t_{i-1} < \tilde t_n <\tilde t_i< \cdots<\tilde t_{n-1}\,,
\end{align*}
so variables from $\tilde t_1$ to $\tilde t_{n-1}$ maintain the same relative order as the original variables, but the variable $\tilde t_n$ is located in the interval $[\tilde t_{i-1},\tilde t_i]$.

Similarly, we can define $n$ changes of variables for the $s_a$ variables
\begin{equation}
s_a =
\left\{
\begin{array}{lllc}
\tilde s_n - \tilde s_{i-a}& & {\rm if}& 1\leq a\leq i-1
\\
\tilde s_n & & {\rm if}&  a= i
\\
\tilde s_n - \tilde s_{n+i-a} + 2 \pi & & {\rm if}& a > i
\end{array}
\right.
\label{changes}
\end{equation}

We now consider each term  in ${\cal W}_n$, labelled by $j$, and change the coordinates $t_a$ to $\tilde t_a$ using (\ref{changet}) for $i=n$ and $s_a$ to $\tilde s_a$ using (\ref{changes}) for $i=j$. For the $n$ terms we obtain the same integrands but different domains of integration for the variables $\tilde s_a$. The resulting domains of integration are such that, when collecting the $n$ terms together, the variable $\tilde s_n$ is integrated from $0$ to $2\pi$ independently of the others. Thus,
\begin{align}
{\cal W}_n = & \underset{0<\tilde t_1< \tilde t_2<\cdots<\tilde t_{n}<2\pi}{\int\! d\tilde t_1\!\int \!d\tilde t_2\!\cdots\!\int d\tilde t_{n}}
\underset{0<\tilde s_1<\tilde s_2<\cdots <\tilde s_{n-1}<2\pi}{\int \!d\tilde s_1\!\int \!d\tilde s_2\cdots\!\int d\tilde s_{n-1}}\int_0^{2\pi} d\tilde s_{n}\ K(\tilde t_n-\tilde s_{n})
\prod_{a=1}^{n-1} K(\tilde t_n-\tilde s_{n}-\tilde t_a+\tilde s_{a})
\nonumber
\\
&\hspace{0.1cm}\times
W(\tilde t_1) W(\tilde t_2-\tilde t_1)\cdots W(\tilde t_{n-1}-\tilde t_{n-2})W(2\pi-\tilde t_{n-1})
\nonumber\\&  \hspace{0.1cm} \times
W(\tilde s_1) W(\tilde s_2-\tilde s_1)\cdots W(\tilde s_{n-1}-\tilde s_{n-2})W(2\pi-\tilde s_{n-1})
 \,.
  \label{nladder2}
\end{align}

Analogous manipulations allow to obtain similar expressions, now with different domains of integration for the variables $\tilde t_a$. Then, at the expense of a $\tfrac{1}{n}$ factor we get an expression
where the variable $\tilde t_n$ is also integrated from $0$ to $2\pi$ independently of the others.
\begin{align}
{\cal W}_n = &\frac{1}n \underset{0<\tilde t_1< \tilde t_2<\cdots<\tilde t_{n-1}<2\pi}{\int\! d\tilde t_1\!\int \!d\tilde t_2\!\cdots\!\int d\tilde t_{n-1}} \int_0^{2\pi} d\tilde t_{n}
\underset{0<\tilde s_1<\tilde s_2<\cdots <\tilde s_{n-1}<2\pi}{\int \!d\tilde s_1\!\int \!d\tilde s_2\cdots\!\int d\tilde s_{n-1}}\int_0^{2\pi} d\tilde s_{n}\ K(\tilde t_n-\tilde s_{n})
\nonumber
\\
&\hspace{0.1cm}\times \prod_{a=1}^{n-1} K(\tilde t_n-\tilde s_{n}-\tilde t_a+\tilde s_{a})
W(\tilde t_1) W(\tilde t_2-\tilde t_1)\cdots W(\tilde t_{n-1}-\tilde t_{n-2})W(2\pi-\tilde t_{n-1})
\nonumber\\&  \hspace{0.1cm} \times
W(\tilde s_1) W(\tilde s_2-\tilde s_1)\cdots W(\tilde s_{n-1}-\tilde s_{n-2})W(2\pi-\tilde s_{n-1})
 \,,
  \label{nladder3}
\end{align}

If we striped off the factor $\frac{1}n$
\begin{equation}
{\cal W}_n = \frac{1}n \tilde{\cal W}_n\,,
\end{equation}
the $\tilde{\cal W}_n$ would correspond to
 the iterative approximations to the Dyson equation written in \cite{Zarembo2001-breaking}. More precisely,
\be
\tilde{\cal W} = \frac{\lambda}{8\pi^2} \int_0^{2\pi} dt \int_0^{2\pi} ds \ \Gamma(2\pi,2\pi;t-s) K(s-t)\,,
\label{def:Zarembo correlator}
\ee
where
\begin{equation}
		\Gamma(s,t;\varphi) = W(s)W(t) + \frac{\lambda}{8\pi^2} \int_0^{s}ds' \int_0^t dt' W(s-s')W(t-t') K(s'-t'+\varphi) \Gamma(s',t';\varphi)\,,
		\label{def:BS}
\end{equation}
with the given boundary condition
	\begin{equation}
		\Gamma(0,t;\varphi) = \Gamma(s,0;\varphi) = 1\,.
		\label{def:BC}
	\end{equation}

 Because of the discrepancy between $\tilde{\cal W}_n$ and ${\cal W}_n$, the quantity $\tilde{\cal W}$ is not the connected correlator. However, $\tilde{\cal W}$ can be easily related to it. Adding a factor $\varepsilon^n$ to every ${\cal W}_n$ we can define
\be
{\cal W}(\varepsilon) =\sum_{n=1}^\infty
\left(\frac{\lambda}{8\pi^2}\right)^n \varepsilon^n {\cal W}_n\,,
\ee
which can be given by the solution of the Dyson equation (\ref{def:Zarembo correlator})-(\ref{def:BC}) through
\be
{\cal W}(\varepsilon) = \int_0^\varepsilon  \frac{\tilde {\cal W}(\varepsilon')}{\varepsilon'}\ d\varepsilon'\,.	
\label{integralrelation}
\ee

Eventually, to get the connected correlator we set $\varepsilon = 1$
\be
\langle W (\mathcal{C}_1,\mathcal{C}_2)\rangle_c = {\cal W}(1)\,.
\ee

The advantage of $\tilde{\cal W}$ over ${\cal W}$ is that the Dyson problem (\ref{def:Zarembo correlator})-(\ref{def:BC}) can be  mapped to a Schr\"odinger-like equation as shown in \cite{Zarembo2001-breaking} and reviewed in the appendix. In the strong-coupling limit the Schr\"{o}dinger problem further simplifies due to the rapidly oscillating behavior of the energy eigenvalues and the corresponding expectation value is saturated by the singularities of the energy distribution thus obtaining
\begin{equation}
		\tilde{\cal W}(\varepsilon)
		\simeq
		e^{2\sqrt{\lambda}\omega_0 (\varepsilon)}\,,
		\label{eq:ev_wl_zarembo}
\end{equation}
where $\omega_0(\varepsilon)$ is the singularity of the integrand (\ref{W}) with largest real part. In the strong coupling limit this singularity is either  $\omega_0(\varepsilon) =1 $, that originates from a square root branch point in the rainbow diagrams, or the value of $\omega$ for which
the ground state of the operator $\hat H_\omega$ (\ref{Homega}) vanishes, which  can be obtained by solving the following equation \cite{Zarembo2001-breaking}
\begin{equation}
E_0(\omega) \approx \frac{1}{4} \left(\omega + \sqrt{\omega^2-1}\right)^2 - \frac{\varepsilon}{2} K(\varphi_{min}) = 0
\end{equation}
where  $- \frac{\varepsilon}{2} K(\varphi) = V(\varphi)$ plays the role of the potential in the
Schr\"{o}dinger problem and has to be evaluated at its minimum $\varphi_{min}$, which of course will depend on the internal space separation $\gamma$.  Equation (\ref{eq:ev_wl_zarembo}) shows the large $\lambda$ behavior we have expected from a string theory computation. In the strong coupling limit, the function ${\cal W}(\varepsilon)$ should behaves as $e^{2\sqrt{\lambda} v (\varepsilon)}$ so, from the derivative of the relation (\ref{integralrelation}) with  respect to $\varepsilon$,
\begin{equation}
e^{2\sqrt{\lambda} v (\varepsilon)} 2\sqrt{\lambda} v'(\varepsilon) \simeq \frac{1}{\varepsilon} e^{2\sqrt{\lambda} \omega_0(\varepsilon)}\,.\nonumber
\end{equation}
Therefore, to leading order in the large $\lambda$ limit, $e^{2\sqrt{\lambda} v (\varepsilon)}\simeq e^{2\sqrt{\lambda} \omega_0(\varepsilon)}$ and there is no difference between $\tilde{\cal W}$ and ${\cal W}$.

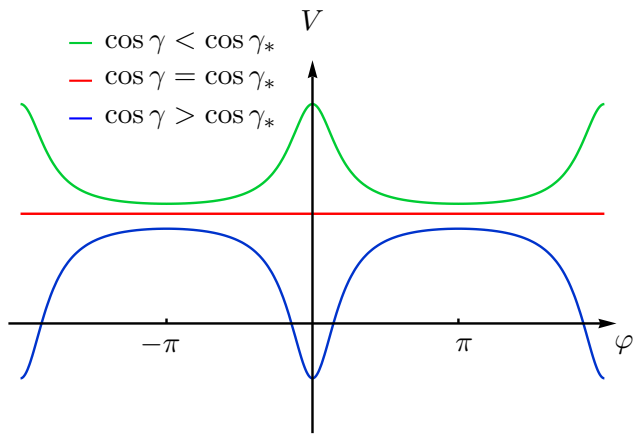
\begin{figure}[h]
\centering
\begin{tikzpicture}
\node at (0,0) {\includegraphics[width=8cm]{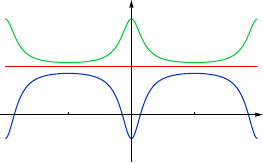}};
\node at (0,3) {$V$};
\node at (-2,-1.3) {$-\pi$};
\node at (2,-1.3) {$\pi$};
\node at (4.1,-1.3) {$\varphi$};
\draw[thick,blue4] (-3.2,2.7)--(-2.9,2.7);
\node at (-1.6,2.7) {$\cos\gamma<\cos\gamma_*$};
\draw[thick,red] (-3.2,2.2)--(-2.9,2.2);
\node at (-1.6,2.2) {$\cos\gamma=\cos\gamma_*$};
\draw[thick,blue0] (-3.2,1.7)--(-2.9,1.7);
\node at (-1.6,1.7) {$\cos\gamma>\cos\gamma_*$};
\end{tikzpicture}
\caption{Form of the Schr\"{o}dinger potential for different relations between $\cos \gamma$ and $\cos \gamma_*$. The value of $\varphi_{min}$ shifts from $0$ to $\pi$ depending on the aforementioned relation.}
\label{potentials}
\end{figure}

When it comes to identify the minimum of the potential, a critical relation is observed between the internal space separation $\gamma$ and spatial separation $h$. If we define
\begin{equation}
\cos \gamma_* = -\left(\frac{h^2+R_1^2+R_2^2}{2R_1R_2}\right)\,,
\label{constraint}
\end{equation}
it is easy to see, as depicted in figure \ref{potentials}, that the minimum  of the potential $V$ is at $\varphi_{min} =0$ when $\cos\gamma>\cos\gamma_*$ and at $\varphi_{min} =\pi$ when $\cos\gamma<\cos\gamma_*$. For the critical value $\cos \gamma = \cos \gamma_*$ the potential becomes constant and the resummation of ladders reduces to a combinatorial problem. In this critical case the contribution of ladder diagrams can be computed exactly as a function of the 't Hooft coupling $\lambda$, as we do in the next section.

For real values of the parameters $\gamma$ and $h$ we are always in the case  $\cos\gamma>\cos\gamma_*$,
so the vanishing of the minimum energy takes the form
\begin{equation}
\frac{1}{4} \left(\omega +\sqrt{\omega ^2-1}\right)^2 +
\frac{1}{4} \frac{1+\cos\gamma}{1+\cos\gamma_*}=0\,.
\label{eq:w0}
\end{equation}
Since $\cos\gamma_* < -1$ in this case, the second term in (\ref{eq:w0}) is negative, but to have a real solution it has to be smaller than $-1/4$. Then, the solution to (\ref{eq:w0}) is real for $\cos\gamma > - \cos\gamma_*-2$ and imaginary for $\cos\gamma < - \cos\gamma_*-2$. In the 
latter case, the singularity with largest real part is 1. Therefore,
\begin{equation}
\omega_0=
\left\{
\begin{split}
& \hspace{2cm}1 &\qquad \textnormal{if} &\,  \cos\gamma < - \cos\gamma_*-2
\\
& \frac{\frac{h^2+R_1^2+R_2^2}{2R_1R_2} + \cos\gamma}{2 \sqrt{1+\cos\gamma}\sqrt{\frac{h^2+R_1^2+R_2^2}{2R_1R_2}-1}} &\qquad \textnormal{if} &\,
\cos\gamma > - \cos\gamma_*-2
\end{split}
\right.
\label{resu:w0}
\end{equation}

In analogy with the Gross-Ooguri phase transition, the ladder contribution to the connected correlator
presents a phase transition as well. When
\be
h > h_c = \sqrt{-(R_1-R_2)^2 + 2R_1R_2(1+\cos\gamma) }\,,
\ee
rainbow diagrams dominate over the diagrams connecting the two circles.

\begin{figure}[h]
\centering
\begin{tikzpicture}
\node at (0,0) {\includegraphics[width=8cm]{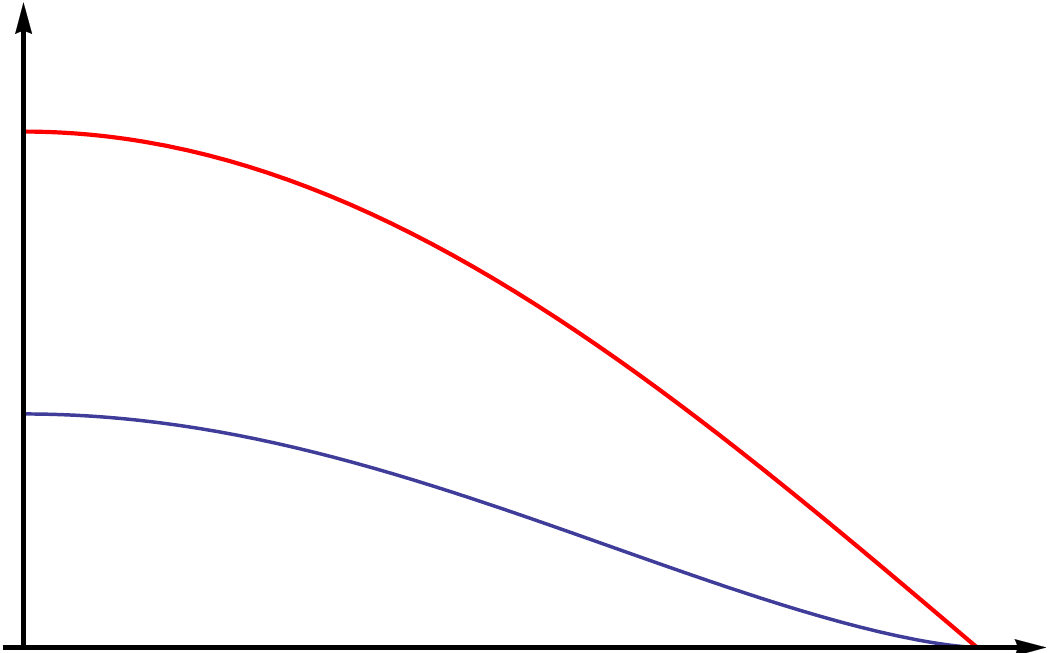}};
\node at (3.4,-2.7) {$\pi$};
\node at (4.,-2.7) {$\gamma$};
\node at (-4.1,2.2) {$h_c$};
\node at (-4.2,1.5) {$2R$};
\draw[thick] (-3.8,1.48)--(-3.7,1.48);
\draw[thick] (3.4,-2.45)--(3.4,-2.35);
\end{tikzpicture}
\caption{Critical separation as a function of the internal space angle $\gamma$ for the case $R_1=R_2=R$. The blue curve represents the transition between connected/disconneted string configurations and the red curve represents the transition in the ladder contribution.}
\label{dostransi}
\end{figure}
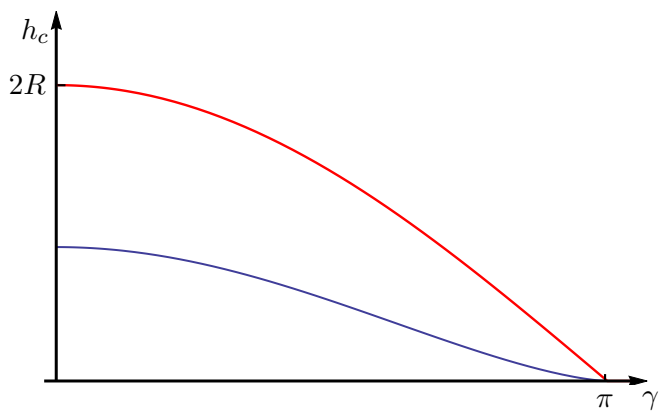

Admittedly, the analogy with the Gross-Ooguri phase transition can only be qualitative, because ladder diagrams account for the connected correlator only partially. In figure \ref{dostransi} we show the critical spatial separation for both phase transitions.

\subsection{Analytical continuation of the ladder resummation}

More interesting is perhaps to consider an analytic continuation of the $\gamma$ parameter such that $\cos \gamma \gg 1$. In this limit $\cos \gamma$ is well above of the critical case. Then, the leading contribution to $\tilde{\cal W}$ is given in terms of the $\omega_0$ specified the second line of (\ref{resu:w0}). Taking $\cos \gamma \gg 1$  we obtain for the ladder contribution to the connected correlator
\begin{equation}
\log \left(
\langle W (\mathcal{C}_1,\mathcal{C}_2)\rangle_c^{\rm ladder} \right)
\simeq \sqrt{\lambda} e^{-\frac{i\gamma}{2}} \sqrt{\frac{R_1 R_2}{h^2 + (R_1-R_2)^2}}\,.
\label{resultado2}
\end{equation}
As we have expected, in this limit ladder diagrams overwhelm the interaction ones and (\ref{resultado2}) precisely agrees with the string theory computation (\ref{resultado1}).

\subsection{Exact resummation of ladders in a critical case}
As we have already seen, there is a critical value of the internal space separation
\begin{equation}
		\cos \gamma = \cos \gamma_*\,,
		\label{constraint2}
\end{equation}
for which the kernel for the ladder diagrams (\ref{def:kernel}) becomes constant
\begin{equation}
	K(\varphi) = -\frac{1}{2}\frac{\cos\gamma + \cos\varphi}{\cos\gamma_* + \cos\varphi} = -\frac{1}{2}\,.
\end{equation}
This constitutes a major simplification, since all propagators connecting ${\cal C}_1$ with ${\cal C}_2$ contribute with a constant factor $-\tfrac{\lambda}{4}$. In addition, rainbow propagators in one circle or the other contributes with a constant factor as well: $\tfrac{\lambda}{4}$. Therefore, the problem of summing all the connected ladder diagrams reduces to a combinatorial problem, which we would like to solve in the planar limit. The  ladder contribution to the connected correlator can then be expanded as
\begin{equation}
\langle W({\cal C}_1,{\cal C}_2)\rangle_{c}^{\rm ladder}
=\sum_{n=1}^\infty\left(\frac{\lambda}{4}\right)^n C^{(n)}\,.
\label{eq:sol_expect_cc}
\end{equation}
To a given perturbative order, we have to consider all possible ways of contracting a total number $n$ of propagators in the expansions of $W({\cal C}_1)$ and  $W({\cal C}_2)$. If $i$, the number of ladder propagators -connecting one circle with the other- is odd (even) the contribution is then negative (positive). Thus, it is convenient to split the contribution from $n$ propagators into
\begin{equation}
		C^{(n)} = \sum_{i=1}^n  (-1)^i C_i^{(n)}\,,
		\label{suma_ladders}
	\end{equation}
where $C_i^{(n)}$ is the contribution of diagrams  with a total of $n$ propagators,  with $i$ of them  being ladder propagators. Consider for example a diagram with $j$ rainbow propagators in ${\cal C}_1$ and $n-i-j$ rainbow propagators in ${\cal C}_2$. Such diagram comes from the expansion of the exponential in $W({\cal C}_1)$ to $(2j+i)^{\rm th}$ order and the exponential in $W({\cal C}_2)$ to $(2n-2j-i)^{\rm th}$ order, so they come with the inverse of $(2j+i)!(2n-2j-i)!$ as a factor. In order to count planar diagrams, we only consider graphs where propagators do not cross and where rainbows do not pass over the endpoint of ladder propagators.
To find the coefficients $C_i^{(n)}$ we can think the $i$ ladder propagators as defining $i$ compartments in each circle, where one has to planarly distribute  $n-i$ rainbow propagators.

To organize the computation we will first count in how many inequivalent ways we can take $i$ ladder propagators among the $2j+i$ points in ${\cal C}_1$ and the $2n-2j-i$ points in ${\cal C}_2$. This requires to choose a point in ${\cal C}_1$ and another in ${\cal C}_2$ to determine the first ladder propagator, which brings a factor of $(2j+i)(2n-2j-i)$. That seems to conclude the first part of the counting, because which other points of  ${\cal C}_1$ and  ${\cal C}_2$ would be connected is determined for the number of rainbow propagators in each compartment. However, with this reasoning we would be incurring in some overcounting. For instance, picking a point in ${\cal C}_1$ for the first ladder propagator and filling the compartments with $(k_1,k_2,\cdots k_i)$ rainbow propagators produces the same diagram as if the first ladder propagator was moved $k_1+1$ points and the compartments filling were $(k_2,\cdots k_i,k_1)$. Then, we have to divide by $i$ to eliminate the overcounting due to this kind of cyclic redefinitions\footnote{There is an analogous factor from the cyclic redefinitions of compartments in ${\cal C}_2$ but this is compensated with the factor from the ways of connecting planarly $i$ points in each circle. }.

To complete the computation, now we have to count $S^{(i)}_j$ and $S^{(i)}_{n-i-j}$, the possible ways of distributing planarly $j$ rainbow propagators in the $i$ compartments of ${\cal C}_1$ and $n-i-j$ rainbow propagators in the $i$ compartments of ${\cal C}_2$ respectively. They will give
\begin{equation}
		C_i^{(n)} = \sum_{j=0}^{n-i}
		\frac{1}{(2j+i)!} \frac{1}{(2n-2j-i)!}
		\frac{(2j+i)(2n-2j-i)}{i} S_{n-i-j}^{(i)} S_j^{(i)}\,.
		\label{coeficientes_ci}
\end{equation}

We can express $S^{(i)}_j$ in terms of $A_k$, that counts the number of planar ways of putting $k$ rainbow propagators in a single compartment
\begin{equation}
		S^{(i)}_k = \sum_{j_1=0}^k\sum_{j_2=0}^{k-j_1} \sum_{j_3=0}^{k-j_1-j_2} \cdots \sum_{j_{i-1}=0}^{k-j_1 - \cdots - j_{i-2}} A_{j_1} A_{j_2} \cdots A_{k-j_1 - \cdots  - j_i}\,.
		\label{recursion_definicion}
	\end{equation}
The quantity $A_k$ is the same as the number of planar rainbows diagrams out of $k$ propagators in a circular Wilson loop. This quantity satisfies a recursion relation \cite{erickson_semenoff}
\begin{equation}
		A_{k+1} = \sum_{j=0}^k A_{k-j} A_j \,,
		\label{eq:zarembo_suma}
\end{equation}
that is solved by
\begin{equation}
		A_k = \frac{\left(2k \right)!}{(k+1)!k!} \, .
		\label{eq:zarembo_coef}
\end{equation}

Although numbers $A_{j_i}$ are known, doing the successive finite sums in (\ref{recursion_definicion}) is not straightforward. Instead, $S_{k}^{(i)}$ can be obtained from a recursive relation that
 follows from the definition (\ref{recursion_definicion}) and the relation (\ref{eq:zarembo_suma})
	\begin{equation}
		S_{k}^{(i)} = S_{k+1}^{(i-1)} - S_{k+1}^{(i-2)}\,.
		\label{recursion}
	\end{equation}
Then, solving this recursive equation we get
\begin{equation}
S_{k}^{(i)} = \frac{i(2k+i-1)!}{k!(k+i)!}\,.
\end{equation}

Replacing this in (\ref{coeficientes_ci}) we obtain,
\begin{equation}
C_i^{(n)} = \sum_{j=0}^{n-i}\frac{i}{j!(j+i)!(n-i-j)!(n-j)!}
= \frac{i(2n)!}{(n!)^2(n+i)!(n-i)!}
\end{equation}
and with this
\begin{equation}
C^{(n)} = -\frac{n}{2(2n-1)}\frac{(2n)!}{(n!)^4}\,.
\end{equation}

Now we can perform the sum (\ref{eq:sol_expect_cc}) and therefore obtain the ladder contribution to the connected correlator exactly as a function of $\lambda$, in terms of modified Bessel functions
\begin{equation}
\langle W(\mathcal{C}_1, \mathcal{C}_2) \rangle_{c}^{\rm ladder} = -\frac{1}{2} \left(\lambda  I_0\left(\sqrt{\lambda }\right){}^2-\sqrt{\lambda } I_1\left(\sqrt{\lambda }\right) I_0\left(\sqrt{\lambda }\right)-\lambda  I_1\left(\sqrt{\lambda }\right){}^2\right)\,.
\label{exactcorrelator}
\end{equation}

In the strong coupling limit (\ref{exactcorrelator}) is
\begin{equation}
		\langle W(\mathcal{C}_1, \mathcal{C}_2)\rangle_{c}^{\rm ladder} \simeq e^{2\sqrt{\lambda}}\,,
\end{equation}
which is of course in agreement with (\ref{eq:ev_wl_zarembo}) and (\ref{resu:w0}) and matches the corresponding string theory calculation since the disconnected world-sheet dominates over the connected one.

This critical case is interesting, not only because the ladder contribution can be exactly computed, but also because it is possible to argue that (\ref{exactcorrelator}) exactly accounts for the connected correlator  in the planar limit.
The reason for that is the fact that the correlator of Wilson loops in the critical case is BPS, as we show in what follows.

Exact results for correlators of  Wilson loops operators have been obtained before via a matrix model calculations
\cite{drukker_gross_exact,Pestun:2007rz,Preti:2016hhk,Giombi:2009ds,Bassetto:2009rt,Giombi:2009ms}. Our family of supersymmetric correlators includes the case $R_1 = R_2 = R$, $h = 0$ and $\gamma = \pi$, which has been studied with matrix model techniques.  In particular, \eqref{exactcorrelator} precisely agrees with eq. (8.31)  in \cite{Giombi:2009ms}. To argue that (\ref{exactcorrelator}) can be obtained from the same matrix model, derived with supersymmetric localization arguments, we will show that our critical correlators are supersymmetric, invariant under the same set of supersymmetry transformations that the case studied in \cite{Giombi:2009ms}.

In order to study the supersymmetry condition let us put the four spinors of ${\cal N} =4$ super Yang-Mills into a single ten-dimensional Weyl spinor $\Psi$. The supersymmetry transformations of the bosonic fields are
\begin{equation}
\begin{split}
\delta_\epsilon A_\mu & = \bar\Psi\Gamma_\mu \epsilon\,,
\\
\delta_\epsilon \Phi_i & = \bar\Psi\Gamma_i \epsilon\,,
\end{split}
\end{equation}
where $\Gamma_A = (\Gamma_\mu,\Gamma_i)$ are ten-dimensional Dirac matrices and the transformation parameter
$\epsilon$ is a ten-dimensional Weyl spinor, which can be split into Poincar\'e and superconformal transformations. Indeed,
\be
\epsilon(x)=\epsilon_0 + x^\mu \Gamma_\mu \epsilon_1\,,
\ee
where $\epsilon_0$ and $\epsilon_1$ are constant Weyl spinors with opposite chiralities.

Now, the supersymmetry variations of the Wilson loop (\ref{wilsonloop}) vanishes, provided that
\be
(i \Gamma_\mu \dot{x}^\mu + \Gamma_i n^i|\dot{x}|)\epsilon(x(\tau))=0\,,
\label{susycond}
\ee
In we consider a circular Wilson loop with
\be
x^\mu = (R \cos\tau,\pm R\sin\tau,h,0)\,,\qquad n^i=(\cos\gamma,\sin\gamma,0,0,0,0)\,,
\ee
where $\pm$ stands for the orientation of curve, equation (\ref{susycond}) becomes
\be
\label{eq:susycondcircle}
(- i \Gamma_1 R \sin \tau \pm i \Gamma_2 R\cos\tau + \Gamma_5\cos\gamma +\Gamma_6\sin\gamma)
(\epsilon_0 + R\cos\tau\Gamma_1\epsilon_1\pm R\sin\tau\Gamma_2\epsilon_1 + h\Gamma_3\epsilon_1)=0\, .
\ee
Equation (\ref{eq:susycondcircle}) is satisfied for any value of the curve parameter $\tau$ if
\be
\epsilon_0 = \left[
\pm i R \left(\Gamma_5\cos\gamma + \Gamma_6\sin\gamma \right)\Gamma_{12} - h\Gamma_3
\right]\epsilon_1\,,
\ee

We now consider the two Wilson loops with opposite orientations given by (\ref{def:trayect}). For them to share some fraction of the supersymmetry one has to simultaneously impose
\begin{align}
\epsilon_0  &=  i R_1 \Gamma_5 \Gamma_{12} \epsilon_1\,,
\label{susy1}
\\
\epsilon_0 &= \left[
- i R_2 \left(\Gamma_5\cos\gamma + \Gamma_6\sin\gamma \right)\Gamma_{12} - h\Gamma_3
\right]\epsilon_1\,.
\label{susy2}
\end{align}
By substituting (\ref{susy1}) into (\ref{susy2}) one finds
\be
\left[1+\frac{R_1}{R_2}\left(\cos\gamma-\sin\gamma\Gamma_{56}\right)-i\frac{h}{R_1}\Gamma_{1235}
\right]\epsilon_0=0\,.
\label{twoWLcond}
\ee
Non-trivial solutions to equation (\ref{twoWLcond}) can be found only if the determinant of matrix acting on $\epsilon_0$ vanishes. To see this we can work with a specific representation for the Dirac matrices. If one chooses them to be hermitian, it is possible to adopt a basis in which
\be
\Gamma_{56} = i\mathbb{1}_{16}\otimes\sigma_3\,,\qquad
\Gamma_{1235} = \mathbb{1}_{16}\otimes\sigma_2\,.
\ee
Thus
\begin{align}
\det\left[\mathbb{1}_{32}+\frac{R_1}{R_2}\left(\cos\gamma \mathbb{1}_{32}-\sin\gamma\Gamma_{56}\right)-i\frac{h}{R_1}\Gamma_{1235}
\right]
&= {\rm det}^{16}
\left(
\begin{array}{cc}
1 + \frac{R_2}{R_1}e^{-i\gamma} & -\frac{h}{R_1}
\\
+\frac{h}{R_1} & 1 + \frac{R_2}{R_1}e^{i\gamma}
\end{array}\right)
\\
&= \left(\frac{h^2+R_1^2+R_2^2 + 2\cos\gamma R_1 R_2}{R_1^{2}}\right)^{16}\,.
\end{align}
Therefore, the two Wilson loops have common supersymmetries only
when\footnote{The critical relation is satisfied for the concentric circles (1.7) considered in \cite{Pestun:2007rz}, which are invariant under the same supersymmetry generator.}
\be
\cos\gamma = - \frac{R_1^2+R_2^2+h^2}{2R_1R_2}\,,
\ee
that corresponds to the constant potential (\ref{constraint}) previously discussed.

\section{Discussion}

We have further developed the study of connected correlators between concentric circular Wilson loops, by taking them with different internal space orientations. We have computed the correlators in the strong coupling limit by calculating the regularized minimal area of world-sheets stretching between concentric circles and using the AdS/CFT correspondence.

From a perturbative point of view we have focused in certain type of Feynman diagrams known as ladder diagrams since its resummation can be related to a Schr\"odinger problem, which can be approximatively solved in the large 't Hooft coupling limit. In general the strong coupling limit of the ladder diagrams resummation does not match the string theory computation because one is not taking into account the contribution of interaction diagrams. Nevertheless, there is a qualitative matching since the ladders resummation presents a phase transition that resembles the Gross-Ooguri phase transition.

Then, we have shown that matching between string theory and ladder computations can also be quantitative when we consider certain analytic continuation of the internal space separation $\gamma$. One of our main results is the matching of the string theory computation (\ref{resultado1}) with the ladder resummation (\ref{resultado2}) when $\cos\gamma\gg1$. This is understood by the fact that at every perturbative order the weight of ladder diagrams overwhelms the weight of interaction diagrams.

Finally we have considered in detail some critical value for the internal space separation $\cos\gamma = - \frac{R_1^2+R_2^2+h^2}{2R_1R_2}$. In that critical case the ladder contribution can be exactly resummed and its strong coupling limit also matches the string theory computation. In this case, the agreement should be presumably explained by a cancellation of interaction diagrams. As shown at the end of our paper the correlator of Wilson loops becomes supersymmetric for the critical internal space separation.

Concerning this case, it would be interesting if the 2-loop perturbative computation of \cite{Plefka2001} could be generalized by an internal space separation between the Wilson loops and verify if the interaction diagrams cancel for the critical value.

\section*{Acknowledgements}
We are thankful to K. Zarembo for useful correspondence. DHC and PP are supported by CONICET.  DHC is also supported by grants PICT 2012-0417, PIP 0681 and PI {\it B\'usqueda de nueva F\'\i sica}.

\appendix
\section{Schr\"odinger problem from the Dyson equation}

We denote by $W(s)$ the whole contribution of all rainbow diagrams within a circular arc of length $s$, and by $W(z)$ its Laplace transform,
\begin{align}
  W(z)=\int_0^\infty ds\ e^{-zs}\,W(s)\,,\qquad
  W(s)=\frac{1}{2\pi i}\int_{c-i\infty}^{c+i\infty} dz\ e^{zs}\,W(z)\,,
\end{align}
where $c\in\mathbb{R}$ is larger than the real parts of all the singularities of $W(z)$. Since the propagator between two points in the circular arc is constant the result can be computed in terms of a Gaussian matrix model and reads
\begin{align}\label{wes}
  W(z)=\frac{8\pi^2}{\lambda}\left( z-\sqrt{z^2-\frac{\lambda}{4\pi^2}}\right)\,,\qquad
  W(s)=\frac{4\pi}{\sqrt{\lambda}\,s}\,I_1\left(\frac{\sqrt\lambda\,s}{2\pi}\right)\,.
\end{align}
It is important to note that $W(z)$ has a branch cut at $z=\pm\sqrt\lambda/2\pi$.

Let us now consider the Dyson equation proposed in \cite{Zarembo2001-breaking},
\begin{align}
  \Gamma(s,t;\varphi)=W(s)W(t)+\frac\lambda{8\pi^2}\int_0^s ds'\int_0^t dt'\ W(s-s')W(t-t')\,K(\varphi+s'-t')\,\Gamma(s',t';\varphi)\,,
  \label{Dysoneq}
\end{align}
where $K(\varphi)$ is the propagator connecting two points in different arcs with phase difference $\varphi$. Upon Laplace transformation in both variables $s,t$ this equation reads
\begin{align}
  \Gamma(z,w;\varphi)=W(z)W(w)
  \left\{1+\frac\lambda{8\pi^2}\sum_{n\in\mathbb{Z}}K_n\,e^{in\varphi}\,\Gamma(z-in,w+in;\varphi)\right\}\,,
\end{align}
where $K_n$ represent the Fourier coefficients of the periodic function $K(\varphi)$. It is now convenient to perform the change of variables
\begin{align}
  z=\omega+ip\,,\qquad
  w=\omega-ip\,,
\end{align}
and to define
\begin{align}
  L(\omega,p;\varphi)=e^{\frac12(w-z)\varphi}\,\Gamma(z,w;\varphi)\,.
\end{align}
Dyson equation (\ref{Dysoneq}) can then be written as
\begin{align}
  \frac{L(\omega,p;\varphi)}{W(\omega+ip)W(\omega-ip)}
  -\frac{\lambda}{8\pi^2}\sum_{n\in\mathbb{Z}}K_n\,L(\omega,p-n;\varphi)=e^{-i p\varphi}\,.
\end{align}
If we introduce the Fourier transformation
\begin{align}
  L(\omega,\theta;\varphi)=\frac{1}{2\pi}\int_{-\infty}^\infty dp\ e^{i p\theta}\,L(\omega,p;\varphi)
\end{align}
we obtain
\begin{align}\label{equ}
  \hat H_\omega\,L(\omega,\theta;\varphi)=\delta(\theta-\varphi)\,,
\end{align}
where $\hat H_\omega$ is the operator
\begin{align}
  \hat H_\omega=\frac{1}{W(\omega+i\hat{p})W(\omega-i\hat{p})}-\frac{\lambda}{8\pi^2}\,K(\theta)
  \label{Homega}
\end{align}
and $\hat{p}$ represents the differential operator $-i\partial_\theta$. The solution to equation \eqref{equ} can then be written as
\begin{align}
  L(\omega,\theta;\varphi)=\sum_{E_\omega}\,\frac{1}{E_\omega}\,\psi_{E_\omega}(\theta)\psi_{E_\omega}^*(\varphi)\,,
\end{align}
where $E_\omega,\psi_{E_\omega(\theta)}$ are the eigenvalues and eigenfunctions of the operator $\hat H_\omega$. Finally, note that
$L(\omega,\varphi;\varphi)$ and $\Gamma(s,s;\varphi)$ are related by a Laplace transformation, such that
\begin{align}
  L(\omega,\varphi;\varphi)
  =\int_0^\infty ds\ e^{-2\omega s}\,\Gamma(s,s;\varphi)\,.
\end{align}

In conclusion,
\begin{align}\label{W}
  \tilde{\mathcal W} &= {\frac{\lambda}{8\pi^2}\int_0^{2\pi} ds \int_0^{2\pi} dt\ K(s-t)\,\Gamma(2\pi,2\pi;s-t)} \nonumber\\
  &=\frac{\lambda}{4\pi}\,\int_{c-i\infty}^{c+i\infty}\frac{d\omega}{2\pi i}\ e^{4\pi\omega}
  \ \sum_{E_\omega}\,\frac{1}{E_\omega}\int_0^{2\pi} d\varphi\ K(\varphi)\,|\psi_{E_\omega}(\varphi)|^2\,,
\end{align}
where $c\in\mathbb{R}$ is larger than the real parts of all the singularities of $L(\omega,\varphi;\varphi)$ in the $\omega$-complex plane. As a matter of fact, the integral in $\omega$ collects the contributions of the singularities of the integrand, which stem from the branch cut at $\omega=\pm\sqrt\lambda/2\pi$ as well as from those values of $\omega$ for which $\hat{H}_\omega$ admits a zero mode $E_\omega=0$. Due to the exponential factor, the leading behavior of $\tilde{\mathcal W}$ in the strong coupling limit is given by the largest of these singular values of $\omega$.

Since there is a branch cut at $\omega=\pm\sqrt\lambda/2\pi$, in order to study these singularities in the strong coupling limit it is convenient to rescale $\omega\to\sqrt\lambda \omega/2\pi$; the operators $W^{-1}(\omega\pm i\hat p)$ then read
\begin{align}
  \frac{1}{W(\sqrt\lambda \tfrac{\omega}{2\pi}\pm i\hat p)}=\frac{\sqrt\lambda}{4\pi}
  \left[\left(\omega\pm 2\pi i\frac{\hat p}{\sqrt\lambda}\right)
  +\sqrt{\left(\omega\pm 2\pi i\frac{\hat p}{\sqrt\lambda}\right)^2-1}\right]\,.
\end{align}
Thus, for $\lambda\gg 1$ the kinetic term $\hat p$ becomes irrelevant. Moreover, the largest value of $\omega$ for which $\hat H_\omega$ has a zero mode corresponds to an operator whose lower bound vanishes,
\begin{align}\label{secsing}
  E_0(\omega)=\frac{\lambda}{4\pi^2}
  \left\{\frac14\left(\omega+\sqrt{\omega^2-1}\right)^2-\frac12K(\varphi_{\rm min})\right\}=0\,.
\end{align}
In this expression $\varphi_{\rm min}$ minimizes the potential $V = -\frac{1}{2}K(\varphi)$ in $\varphi\in[0,2\pi)$. In consequence, the leading contribution to $\tilde{\cal W}$ is
\begin{align}
\tilde{\cal W} \sim e^{2\omega_0\sqrt\lambda}\,,
\end{align}
where $\omega_0$ is either $1$ or the solution of \eqref{secsing}, depending on whether
$K(\varphi_{\rm min})$ is smaller or larger than $1/2$, respectively.

\end{document}

%% file: plavsnonpla.pdf_tex
\begingroup%
  \makeatletter%
  \providecommand\color[2][]{%
    \errmessage{(Inkscape) Color is used for the text in Inkscape, but the package 'color.sty' is not loaded}%
    \renewcommand\color[2][]{}%
  }%
  \providecommand\transparent[1]{%
    \errmessage{(Inkscape) Transparency is used (non-zero) for the text in Inkscape, but the package 'transparent.sty' is not loaded}%
    \renewcommand\transparent[1]{}%
  }%
  \providecommand\rotatebox[2]{#2}%
  \ifx\svgwidth\undefined%
    \setlength{\unitlength}{379.98535408bp}%
    \ifx\svgscale\undefined%
      \relax%
    \else%
      \setlength{\unitlength}{\unitlength * \real{\svgscale}}%
    \fi%
  \else%
    \setlength{\unitlength}{\svgwidth}%
  \fi%
  \global\let\svgwidth\undefined%
  \global\let\svgscale\undefined%
  \makeatother%
  \begin{picture}(1,0.40336061)%
    \put(0,0){\includegraphics[width=\unitlength,page=1]{plavsnonpla.pdf}}%
    \put(0.12477445,0.00756372){\color[rgb]{0,0,0}\makebox(0,0)[lb]{\smash{planar}}}%
    \put(0.69321643,0.00756372){\color[rgb]{0,0,0}\makebox(0,0)[lb]{\smash{non-planar}}}%
  \end{picture}%
\endgroup%